\documentstyle[12pt]{article}
\begin{document}
\baselinestretch{1.5}

\begin{titlepage}

\begin{flushright}
IASSNS-HEP-92/50
\end{flushright}
\begin{flushright}
IFF-FM-92/7
\end{flushright}
\begin{flushright}
August 1992
\end{flushright}
\begin{center}

\vspace{.5 true  cm}

\Large
{\bf The Electronic Spectrum of Fullerenes}\\
\vspace{6pt}
{\bf from the Dirac Equation}\\

\vspace{1.5 true  cm}

\large
{\bf J. Gonz\'alez\S\footnote{Email: emgonzalez@iem.csic.es},
F. Guinea\footnote{ On leave from: Instituto de Ciencias de Materiales,
CSIC, Cantoblanco, 28050 Madrid, Spain.}\dag and
M.A.H. Vozmediano\ddag\footnote{Address after October 1992:
Instituto de F\'{\i}sica Fundamental, CSIC, 28006 Madrid, Spain.}}\\

\vspace{0.5 true  cm}

\normalsize
\S{\em Instituto deF\'{\i}sica Fundamental, CSIC} \\
{\em Serrano 123, 28006 Madrid, Spain} \\

\dag{\em  The Harrison M. Randall Laboratory of Physics } \\
{\em The University of Michigan, Ann Arbor, Michigan 48109, USA} \\

\ddag{\em The Institute for Advanced Study} \\
{\em Princeton, NJ 08540, USA}

\vspace{.5 true  cm}

\end{center}

\begin{abstract}

The electronic spectrum of sheets of graphite (plane honeycomb
lattice) folded into regular polihedra is studied. A continuum
limit valid for sufficiently large molecules and
based on the tight binding approximation is derived. It is
found that a Dirac
equation describes the flat grafite lattice.
Curving the lattice by insertion of odd
numbered rings can be mimicked by coupling effective gauge fields.
In particular the $C_{60}$ and related molecules are well described
by the Dirac equation on the surface of a sphere coupled to a
color monopole sitting at its center.

\vspace{3 true cm}

\end{abstract}

\vspace{3 true cm}
\vskip5cm
\noindent

\noindent

\eject

\newpage

\end{titlepage}
\section{Introduction}

Fullerene molecules \cite{bol} are carbon cages which appear
in the vaporization of graphite. They have become a source of great fun for
physicist in different areas due to their curious properties and the
relative ease in which they can now be synthesized and manipulated.
One of their most beautiful features from a formal point of view
is their geometrical character and the exciting possibility of
producing them in all sorts of geometrical shapes having as building
blocks sections of the honeycomb graphite lattice.
The more aboundant of them, the $C_{60}$ molecule nicknamed ``bucky ball",
is also the most spherical.
The sixty carbon atoms are placed at the vertices of a truncated
icosahedron, obtained after replacing each of the icosahedron
vertices by a regular pentagon. The shape of the $C_{60}$
molecule is then that of a soccer ball, consisting of 12 pentagons
and 20 hexagons. This molecule has shown to possess striking
magnetic and electric properties, the most important of which is
the superconducting nature of its alkaline compounds.
In the process
of graphite vaporization, there appear together with $C_{60}$
a full set of family members of various categories. Some
are slightly deformed, as the $C_{70}$ whose shape is more
like an american football ---an elliptical deformation--- ,
some others are
directly related with $C_{60}$ in the sense that they are bigger but
have the symmetry group of the icosahedron.
These larger molecules may be imagined as
built from triangular pieces of the honeycomb lattice of the
type shown in figure 1. When the triangles are assembled as the
faces of the icosahedron we end up with a lattice which has constant
coordination number for all the sites. The first molecules of
this series from the $C_{60}$, namely the $C_{240}$ and the
$C_{540}$ molecules, have already been synthesized\cite{ca}.
The honeycomb lattice is very interesting in the study of two dimensional
statistical models as a rather non
trivial tessellation of the plane. The combination of the honeycomb lattice
with the truncated icosahedron structure together with the existence of
larger molecules grown from the mother $C_{60}$ opens a totally  new
field of research, namely this of two dimensional statistical models
on the curved lattices. Moreover, the compact geometry of the
icosahedron is not the only possibility to fold the honeycomb
lattice. The triangular pieces of the type shown in figure 1 can be
matched to form a lattice with constant coordination number
inscribed in the tetrahedron or the
octahedron. It turns out that,
for all regular polyhedra whose faces are equilateral triangles,
one may devise a recursive procedure to build respective series
of lattices with growing number of points and preserving the
symmetry group of the original polyhedra.

The main purpose of the present article is to study the
electronic properties of the series of molecules which, starting
from $C_{60}$, have the symmetry group of the icosahedron. Of
the four valence electrons which has each carbon atom, three of
them build up the $\sigma$ orbitals along the lattice links and
are responsible for the elastic properties. The
remaining $\pi$ orbitals give rise to the conducting properties
of the molecules, which can be considered peculiar lattices
at half-filling. In the description of the electronic
excitations we will rely on the tight-binding method, which
translates the problem into that of finding the spectrum of the hopping
hamiltonian for the fermion operators $a_{i}, a_{j}^{+}$
\begin{equation}
H = \gamma \sum_{<i,j>} a_{i}^{+}a_{j}  \label{0}
\end{equation}
the sum running over nearest neighbors $i,j$ in the curved lattice.
In order to make contact with experiment, two different $\gamma$ hopping
parameters should be taken for pentagon links and for links belonging
to pairs of hexagons. This is consistent with the
fact that in the $C_{60}$ molecule, for instance, the
interatomic distance along a pentagon link is slightly greater than
the distance between neighboring atoms in different
pentagons. It has been pointed out, however, that any ratio of the two
hopping parameters between 0.9 and 1.0 gives quite reasonable
results\cite{spec}. It is important to stress that, in spite of
neglecting the coulombic interaction between the electrons, the
tight-binding approximation predicts energy levels which are in
good quantitive agreement with the existing experimental
results\cite{spec}.

In this paper we undertake the problem of diagonalizing (\ref{0})
in two steps. We first investigate the plane honeycomb lattice
and we then fold it to form the given polyhedron. The pure graphite sheet
is already very interesting in its own. The more important characteristic
we find is that, instead of having a Fermi line, it has
a finite set of isolated Fermi points when
studied at half-filling. This is the physically interesting situation for
carbon lattices as discussed in section 2.  This is the basis
which allows us to build a local field theory to describe the
low energy excitations of the electronic hamiltonian about each
of the Fermi points. The geometry and coordination of the lattice
---it is in fact made of two interpenetrating triangular sublattices---
determines the field theory to be that of a two-component Dirac
spinor, while
the existence of two {\it independent} Fermi points doubles the spectrum.

Next we come to the study of the folded lattices, having in mind in
particular the lattices of the three basic regular polyhedra: tetrahedron,
octahedron and icosahedron. All of them produce by simple truncation
a lattice that can then be grown indefinitely the same way
as the $C_{60}$ does. Being regular polyhedra, all of them can be
flattened off on the plane in a certain way so that we are back to
the study of the honeycomb lattice in a portion of the plane with
rather nontrivial boundary conditions. Notice that this approach,
by emphasizing the translational invariance of the problem, is
orthogonal to the ones that use group theoretical methods based
on the rotational symmetry of the molecule\cite{samuel}.
When this procedure is applied to the tetrahedron the first thing
that we observe is that its conical singularities (vertices),
although truncated, generate lines of frustration in the lattice.
More concretely, its two triangular sublattices are interchanged
by the boundary conditions. As a consequence of it,
we learn about the interchange of Fermi points (or of the
atoms in the basis of the Bravais lattice) that takes place in the
tetrahedron as well as in the icosahedron\footnote{We note
by passing that such
interchange does not take place in the octahedron which will be
treated in a different publication}.
The effect of frustration is then
twofold, since on one side we have curvature and on the other side we have
to couple the formerly independent Dirac spinors to produce the
mentioned interchange. For that we treat the two Dirac
spinors as the two components of an SU(2) color doublet. Each vertex
is traversed by a quantum of color magnetic flux mimicking the
interchange of color amplitudes induced by the vertices of the
polyhedron. We are able in this way to make contact with the
phenomenological study presented in ref. \cite{nos}. There
the effect of the magnetic field was smoothed over the sphere by
considering a monopole sitting at its center, and the
simplified model for the icosahedral
$C_{60}$ with an abelian monopole of charge 3/2
reproduced the observed low-energy spectrum. The refinement
presented here represents a qualitatively better understanding of
the problem and allows to predict the correct value of the
monopole charge.

The presentation of the article goes as follows: in section 2 we study
in full detail the plane honeycomb lattice, perform the tight-binding
approximation and extract the continuum limit. In section 3 we describe
the technique
for solving the free field hamiltonian in the honeycomb lattice folded on
regular polyhedra. The case of the grown tetrahedron with 48
lattice points
is worked out as an explicit example. In section 4 we explain in detail
the phenomenological model proposed to compute the spectrum
of large icosahedral molecules. We will justify the validity of the continuum
limit for the fullerene molecules, discuss the role played by the
peculiar geometry of the honeycomb lattice, and write the appropriate
Dirac equation. In section 5 we will write down a summary of the main
points developed through the paper highlighting the more important
issues and will discuss its implications and prospects.

\section{The planar honeycomb lattice}

We will see in what follows that the tight-binding approximation
applied to the computation of the electronic spectrum in
periodic potentials leads to the equivalent problem of the
spectrum of a free field theory on a lattice. We will focus on
the resolution of the honeycomb lattice on the two-dimensional plane.

In solid state applications one is interested on periodic
potentials which correspond to a given disposition of the atoms
in a crystal. The set of discrete translations under which the
potential $V({\bf r})$ is invariant can be generated by
independent transformations ${\bf T}_{1}$ and ${\bf T}_{2}$ (in
two dimensions) so that
\begin{equation}
V({\bf r} + p{\bf T}_{1} + q{\bf T}_{2}) = V({\bf r}) \;\;\;\;\;\forall
p,q \in Z
\end{equation}
The statement of Bloch theorem is that the energy eigenfunctions
of the quantum problem can be expressed as
\begin{equation}
\Psi_{n {\bf k}}({\bf r}) = \mbox{\Large $e^{i{\bf k \cdot r}}$
} u_{n {\bf k}}({\bf r})   \label{1}
\end{equation}
with $ u_{n {\bf k}}({\bf r})$ being a periodic function
\begin{equation}
 u_{n {\bf k}}({\bf r} + p{\bf T}_{1} + q{\bf T}_{2}) =
  u_{n {\bf k}}({\bf r})  \;\;\;\;\; \forall p,q \in Z
\end{equation}
In the above expression $n$ stands for the band label, while ${\bf
k}$ labels different states in a band. A way of exploiting the
content of Bloch theorem is to write the energy eigenfunctions
(\ref{1}) in the form of the so-called Wannier functions
\begin{equation}
\Psi_{n {\bf k}}({\bf r}) = \frac{1}{\sqrt{N}} \sum_{i}
\mbox{\Large $e^{i{\bf k \cdot r}_{i}}$
} \phi_{n}({\bf r} - {\bf r}_{i})     \label{2}
\end{equation}
where the sum runs over all the lattice points
\begin{equation}
{\bf r}_{i} = p_{i}{\bf T}_{1} + q_{i}{\bf T}_{2}
\;\;\;\;\;\;p_{i},q_{i} \in Z
\end{equation}
It can be shown that (\ref{2}) is a Bloch
wavefunction and, viceversa, that an eigenfunction with the property
(\ref{1}) can always be cast in the form (\ref{2}) \cite{am}.

The simplest instance in applying the tight-binding method
occurs when there is no significant mixing between states
belonging to different bands. Then one can insert one linear
combination (\ref{2}) in the computation of the energy
eigenvalues. Since the energy eigenfunctions are always
expressible in the form (\ref{2}), we may take advantage of the
variational approach to conclude that the energy levels are
given by
\begin{equation}
E_{\bf k}   =   \frac{\int d^{2} r \;\overline{\Psi}_{\bf k} H
\Psi_{\bf k}}{\int d^{2} r \;\overline{\Psi}_{\bf k} \Psi_{\bf k}}
     =   \frac{\sum_{i}\mbox{\Large $e^{i{\bf k \cdot r}_{i}}$}
\int d^{2} r \;\overline{\phi}({\bf r}) H \phi({\bf r} - {\bf r}_{i})  }
{\sum_{i}\mbox{\Large $e^{i{\bf k \cdot r}_{i}}$}
\int d^{2} r \;\overline{\phi}({\bf r})  \phi({\bf r} - {\bf
r}_{i}) }
\end{equation}
In physical situations in which the orbitals $\phi$ are localized
around each lattice site, it may be appropriate to approximate
the numerator by integrals involving only nearest neighbor
orbitals, which is the essence of the tight-binding method. In
practice, it is also reasonable to take the norm of the Wannier
wavefunctions as constant, which we set to one by a proper normalization.
This gives the result for the energy
levels in tight-binding approximation
\begin{equation}
E_{\bf k} = \int d^{2} r \;\overline{\phi}({\bf r}) H \phi({\bf r})
+  \sum_{\stackrel{{\scriptstyle nearest}}{{\scriptstyle neighbors}}
}\mbox{\Large $e^{i{\bf k \cdot r}_{i}}$}
\int d^{2} r \;\overline{\phi}({\bf r}) H \phi({\bf r} - {\bf r}_{i})
\end{equation}
where the sum runs over all nearest neighbors $i$ of the origin.

The derivation we have made is appropriate to lattices which
have only an atom per primitive cell, though the extension of
the method to the general case is straightforward. For the
lattice which is the object of our interest, the honeycomb
lattice of graphite shown in figure 2,
there are two atoms in the primitive cell.
We may take as generators of the lattice the vectors
\begin{equation}
{\bf T}_{1} = \sqrt{3} {\bf e}_{x} \;\;\;\;\;\;\;
{\bf T}_{2} = \frac{\sqrt{3}}{2}{\bf e}_{x} + \frac{3}{2} {\bf e}_{y}
\end{equation}
Then we need to place an atom at the origin of the primitive cell and
another displaced at ${\bf d} = {\bf e}_{y}$, for instance, to
produce the honeycomb lattice by repeated application of the
generators. The wavefunctions $\phi$ may be thought of being
composed of two identical orbitals $\phi_{\bullet}$ and
$\phi_{\circ}$ localized respectively around each of the two
mentioned points of the primitive cell. We may exploit the
variational freedom by considering an arbitrary linear combination
of these two orbitals, which we write in the form
\begin{equation}
\phi({\bf r}) = c_{\bullet} \phi_{\bullet}({\bf r}) +
  c_{\circ}\mbox{\Large $e^{i{\bf k \cdot d}}$}
\phi_{\circ}({\bf r} - {\bf d})
\end{equation}
By introducing this expression into the energy functional we
get, in tight-binding approximation,
\begin{eqnarray}
E_{\bf k} & = & \overline{c}_{\bullet} c_{\bullet}
  \int d^{2} r \;\overline{\phi}_{\bullet}({\bf r}) H
\phi_{\bullet}({\bf r})  \;  + \;
\overline{c}_{\circ} c_{\circ}
  \int d^{2} r \;\overline{\phi}_{\circ}({\bf r} - {\bf d}) H
\phi_{\circ}({\bf r} - {\bf d})   \nonumber  \\
   &  &  +  \overline{c}_{\bullet} c_{\circ}
 \sum_{j}\mbox{\Large $e^{i{\bf k \cdot u}_{j}}$}
\int d^{2} r \;\overline{\phi}_{\bullet}({\bf r}) H
  \phi_{\circ}({\bf r} - {\bf u}_{j})  \nonumber   \\
   &  &  + \overline{c}_{\circ} c_{\bullet}
 \sum_{j}\mbox{\Large $e^{i{\bf k \cdot v}_{j}}$}
\int d^{2} r \;\overline{\phi}_{\circ}({\bf r} - {\bf d}) H
  \phi_{\bullet}({\bf r} - {\bf d} - {\bf v}_{j})
\end{eqnarray}
where $\{ {\bf u}_{j} \}$ is a triad of vectors of link length
pointing respectively in the direction of the nearest neighbors
of a $\bullet$ point, and $\{ {\bf v}_{j} \}$ the triad made of
their respective opposites (see figure 2).
We face the ideal situation in which
neither the $\phi_{\bullet}$ nor the $\phi_{\circ}$ orbitals are
directionated over the two-dimensional plane. Then the symmetry of
the problem imposes that
\begin{eqnarray}
  \int d^{2} r \;\overline{\phi}_{\bullet}({\bf r}) H
\phi_{\bullet}({\bf r})  =   \int d^{2} r \;
\overline{\phi}_{\circ}({\bf r} - {\bf d}) H
\phi_{\circ}({\bf r} - {\bf d})   &  =  &  \beta   \\
  \int d^{2} r \;\overline{\phi}_{\bullet}({\bf r}) H
\phi_{\circ}({\bf r} - {\bf u}_{i})  =   \int d^{2} r \;
\overline{\phi}_{\circ}({\bf r} - {\bf d}) H
\phi_{\bullet}({\bf r} - {\bf d} - {\bf v}_{i})   &  =  &  \gamma
 \;\;\;\;\;\;\; \forall i
\end{eqnarray}

The variational problem for the honeycomb lattice becomes, then,
the diagonalization of the quadratic form
\begin{equation}
E_{\bf k} = (\overline{c}_{\bullet}\;\;\overline{c}_{\circ})
     \left(
\begin{array}{cc}
 \beta  & \gamma  \sum_{j}\mbox{\Large $e^{i{\bf k \cdot u}_{j}}$} \\
\gamma \sum_{j}\mbox{\Large $e^{i{\bf k \cdot v}_{j}}$} & \beta
\end{array} \right) \left(
\begin{array}{c}
c_{\bullet}  \\
c_{\circ}
\end{array}    \right) \label{5}
\end{equation}
We will disregard in what follows the diagonal contribution in
(\ref{5}) as long as it is independent of ${\bf k}$. We will come
back later to the band structure of the honeycomb lattice of graphite.

The approximations made by the tight-binding method reduce the
problem, in practice, to that of a set of coupled oscillators on
the lattice. This can be appreciated in the expression of the
energy functional (\ref{5}), in which what matters is
essentially the coordination between the lattice sites. By
application of the tight-binding method what we are doing
actually is diagonalizing the hamiltonian
\begin{equation}
H = \gamma \sum_{<i,j>} a_{i}^{+}a_{j}  \label{6}
\end{equation}
where the sum is over pairs of nearest neighbors atoms $i,j$ on the
lattice and $a_{i}$ , $a_{j}^{+}$ are canonically anticommuting
operators
\begin{equation}
\{ a_{i},a_{j} \} = \{ a_{i}^{+}, a_{j}^{+} \}
 = 0  \;\;\;\;\;\:\;\;\; \{ a_{i}, a_{j}^{+} \} = \delta_{ij}
\label{7}
\end{equation}
In fact, this problem can be solved by a variant of the method
sketched above. We first form the eigenstate of ${\bf T}_{1}$
and ${\bf T}_{2}$
\begin{equation}
\Psi = \sum_{i \;\bullet} c_{\bullet}
\mbox{\Large $e^{i{\bf k \cdot r}_{i}}$} a_{i}^{+} \left| O \right\rangle
+  \sum_{i \;\circ} c_{\circ}
\mbox{\Large $e^{i{\bf k \cdot r}_{i}}$}a_{i}^{+} \left| O \right\rangle
\label{8}
\end{equation}
assigning different coefficients $c_{\bullet}$ and $c_{\circ}$
to black and blank points, respectively, as depicted in figure
2. Under the action of the hamiltonian, however, black points are mapped into
blank points and viceversa. We have, indeed,
\begin{eqnarray}
 H \Psi & = & \gamma \sum_{i \;\bullet}
\sum_{<i,j>} c_{\circ}\mbox{\Large $e^{i
{\bf k \cdot r}_{j}}$} a_{i}^{+} \left| O \right\rangle
\; + \; \gamma \sum_{i \;\circ} \sum_{<i,j>}c_{\bullet}
\mbox{\Large $e^{i
{\bf k \cdot r}_{j}}$} a_{i}^{+} \left| O \right\rangle
  \nonumber    \\
 & = &   \gamma \sum_{j}  \mbox{\Large $e^{i {\bf k \cdot u}_{j}}$}
\sum_{i \;\bullet} c_{\circ}
\mbox{\Large $e^{i {\bf k \cdot r}_{i}}$}
            a_{i}^{+} \left| O \right\rangle
\;  + \;  \gamma \sum_{j}  \mbox{\Large $e^{i {\bf k \cdot v}_{j}}$}
\sum_{i \;\circ}  c_{\bullet}
        \mbox{\Large $e^{i {\bf k \cdot r}_{i}}$}
           a_{i}^{+} \left| O \right\rangle
\end{eqnarray}
It is clear that the state (\ref{8}) is an eigenvector of
$H$ provided that the coefficients $c_{\bullet}$ and $c_{\circ}$
are solutions of the eigenvalue problem
\begin{equation}
     \left(
\begin{array}{cc}
 0  &  \gamma \sum_{j}\mbox{\Large $e^{i{\bf k \cdot u}_{j}}$} \\
\gamma \sum_{j}\mbox{\Large $e^{i{\bf k \cdot v}_{j}}$} &  0
\end{array} \right) \left(
\begin{array}{c}
c_{\bullet}  \\
c_{\circ}
\end{array}    \right) = E_{\bf k}
\left(
\begin{array}{c}
c_{\bullet}  \\
c_{\circ}
\end{array}    \right) \label{9}
\end{equation}
This is nothing but a different expression of the variational
problem (\ref{5}).

{}From (\ref{9}) a straightforward computation gives the band of levels
\begin{equation}
E_{\bf k} =  \pm \gamma \sqrt{1 + 4
cos^{2}\frac{\sqrt{3}}{2} k_{x} +
4 cos \frac{\sqrt{3}}{2}k_{x} \; cos \frac{3}{2}
k_{y} }          \label{12}
\end{equation}
The structure of this band has very striking properties when
considered at half-filling. This is the situation  which has
physical interest, since in the case of graphite each site of
the honeycomb lattice yields one electron to the Fermi sea. Each
level of the band may accomodate two states due to the spin
degeneracy, and the Fermi level turns out to be at the midpoint
of the band, $E_{\bf k} = 0$. Quite amazingly, the honeycomb
lattice at half-filling has six isolated Fermi points, instead
of a whole Fermi line. In the reciprocal lattice generated by
\begin{equation}
{\bf K}_{1} = \frac{2\pi }{\sqrt{3}} {\bf e}_{x} - \frac{2\pi
}{3} {\bf e}_{y}  \;\;\;\;\;\;\;\;  {\bf K}_{2} = \frac{4\pi }{3}
{\bf e}_{y}
\end{equation}
the first Brillouin zone has as many momenta as primitive
cells contains the original lattice. Such primitive cell of the
reciprocal lattice is an hexagon, as shown in figure 3. The
only points which reach the Fermi level are the six vertices of
the hexagon
\begin{eqnarray}
  k_{x} = \pm \frac{4\pi }{3 \sqrt{3}} &   & k_{y} = 0  \nonumber  \\
  k_{x} = \pm \frac{2\pi }{3 \sqrt{3}} &   & k_{y} = \pm
\frac{2\pi }{3}   \label{26}
\end{eqnarray}
It can be checked that these are the only roots of $E_{\bf k} =
0$. The representation of the lower branch of the band in figure
4 illustrates the peculiar form of the Fermi sea. At last, the
independent number of Fermi points is two, since any two momenta
congruent by ${\bf K}_{1}$, ${\bf K}_{2}$ are just different labels
of the same state.

The existence of a finite number of Fermi points at half-filling has
important consequences in the description of the spectrum about
the Fermi level. The low energy excitations can be studied by
taking the continuum limit at any two independent Fermi points.
As long as the number of them is finite,
the outcome is that a simple
field theory suffices to describe the electronic spectrum of
large honeycomb lattices. The continuum limit can be taken
by na\"{\i}ve scaling of dimensionful quantities since we are
dealing with a free theory. For this purpose we introduce a
parameter $a$ measuring the link length and expand the $2\times
2$ operator in (\ref{9}) at any of two independent Fermi points.
At the first Fermi point in (\ref{26}), for instance, we have
\begin{equation}
{\bf k} = \frac{4\pi }{3\sqrt{3}} {\bf e}_{x} +
\mbox{\boldmath $\delta k$}  \label{66}
\end{equation}
and
\begin{equation}
{\cal H} \equiv     \left(
\begin{array}{cc}
 0  &  \gamma \sum_{j}\mbox{\Large $e^{ia{\bf k \cdot u}_{j}}$} \\
\gamma \sum_{j}\mbox{\Large $e^{ia{\bf k \cdot v}_{j}}$} &  0
\end{array} \right) \approx - \frac{3}{2} \gamma a
     \left(
\begin{array}{cc}
 0  &  \delta k_{x} + i \delta k_{y} \\
 \delta k_{x} - i \delta k_{y} &  0
\end{array} \right) + O((a\;\delta k)^{2})
\end{equation}
The na\"{\i}ve scaling
\begin{equation}
\lim_{a \rightarrow 0} {\cal H}/a = - \frac{3}{2} \gamma \;
\mbox{\boldmath $\sigma$}^{T} \mbox{\boldmath $\cdot \delta k$}
\end{equation}
dictates the effective hamiltonian in the continuum limit, which
turns out to be the Dirac operator in two dimensions. The same
result is obtained at any of the six points in (\ref{26}).
Given the
existence of two independent Fermi points, we conclude that the
low energy excitations of the honeycomb lattice at half-filling
are described by an effective theory of two two-dimensional
Dirac spinors. This result is at odds with the more
standard continuum approximation to lattice theories
in condensed matter physics, the effective mass theory.
There, a quadratic dispersion relation at high symmetry points
of the Brillouin zone gives rise to an effective
Schr\"odinger equation, with one parameter, the mass,
chosen to reproduce the exact curvature.
Only one dimensional systems, and 3D semiconductors
with the diamond structure and no gap, are known to give rise
to the Dirac equation.
to

\section{Folded honeycomb lattices}

In this section we describe the general method by which the free
fermion theory can be solved on homogeneous curved lattices. By an
homogeneous lattice we mean one in which the coordination number
remains constant for all the vertices. These are honeycomb
lattices inscribed on the tetrahedron, the octahedron or the
icosahedron. To illustrate the method we take the particular
case of a generic honeycomb lattice inscribed on the
tetrahedron. On topological grounds, the tetrahedron is an
orbifold, i.e. a manifold with several singular points. There
exists a particular set of coordinates which maps this orbifold
into a bounded region of the two-dimensional plane, as shown in
figure 5 \cite{des}. When such a
coordinate system is chosen the lattice can be unfolded on the
plane, bearing in mind the appropriate identifications of points.

We want to solve the hamiltonian of coupled fermion oscillators (\ref{7})
\begin{equation}
H = \gamma \sum_{<i,j>} a_{i}^{+}a_{j}  \label{91}
\end{equation}
where the sum is now over pairs of nearest neighbors $i,j$ on the
folded honeycomb lattice.
Suppose that we introduce a would-be eigenvector
\begin{equation}
\Psi = \sum_{i} f_{i} a_{i}^{+} \left| O \right\rangle
\label{93}
\end{equation}
Then, under the action of the hamiltonian we have
\begin{equation}
H \Psi = \gamma \sum_{i} \sum_{<i,j>} f_{j} a_{i}^{+}
\left| O \right\rangle       \label{94}
\end{equation}
where the sum on $j$ is over nearest neighbors of the $i$
vertex. We have a solution of the eigenvalue problem if and only
if the quantity
\begin{equation}
\frac{1}{f_{i}} \sum_{<i,j>} f_{j} = \lambda   \label{95}
\end{equation}
is constant over the lattice. In the particular coordinate
system which maps the tetrahedron on the region of figure 5,
it is not difficult to think of functions which satisfy
(\ref{95}), with $\lambda$ being a constant. Actually, the most general
solution  of this difference equation is given by a combination
of exponential functions. We have to require, however, that the
given combination be singlevalued on the tetrahedron. At this
point, it proves useful to turn for a moment to the boundary
value problem on the continuum.

Suppose that we were looking for singlevalued and differentiable
functions on the region of figure 5, with all pertinent
identifications made. The set of plane waves allows to
build a complet set matching all boundary conditions implied by
the identifications. These are, in units in which the side of
the triangular faces is $L$,
\begin{eqnarray}
&  \Psi (0,y) = \Psi (0, -y) &   \label{961} \\
&  {\bf n \cdot \nabla} \Psi (0, y) =
  - {\bf n \cdot \nabla} \Psi(0, -y)  &    \label{962} \\
&  \Psi \left( \frac{\sqrt{3}}{2}L,\frac{1}{2}L + y \right) =
   \Psi \left( \frac{\sqrt{3}}{2}L,\frac{1}{2}L - y \right) & \label{963} \\
&  {\bf n \cdot \nabla}
      \Psi \left( \frac{\sqrt{3}}{2}L,\frac{1}{2}L + y \right) =
  - {\bf n \cdot \nabla}
      \Psi \left( \frac{\sqrt{3}}{2}L,\frac{1}{2}L - y \right) & \label{964} \\
&  \Psi (x,y) = \Psi (x, y + 2L) &  \label{965}
\end{eqnarray}
where ${\bf n}$ denotes always the normal unit vector pointing
outwards the given boundary. The first two
boundary conditions can be satisfied at once by taking
\begin{equation}
\Psi \sim \;\; cos \;({\bf k \cdot r})
\end{equation}
The requirement of periodicity (\ref{965}) implies that
\begin{equation}
k_{y} = \frac{\pi}{L} n \;\;\;\; n \in Z
\end{equation}
Finally, it is easily seen that (\ref{963}) and (\ref{964}) are
satisfied  provided that
\begin{equation}
k_{x} \frac{\sqrt{3}}{2} + k_{y} \frac{1}{2} = \frac{\pi}{L} m \;\;\; m
\in Z
\end{equation}
which gives the constraint
\begin{equation}
k_{x} = \frac{2}{\sqrt{3}}\frac{\pi}{L} \left( m - \frac{n}{2} \right)
\end{equation}
We have found, therefore,  a complete set $\left\{ \Psi_{(m,n)}
\right\}$, with
\begin{eqnarray}
 &  \Psi_{(m,n)} = cos \;({\bf k}^{(m,n)} {\bf \cdot r})  & \label{97}  \\
 &  {\bf k}^{(m,n)}  = \frac{2}{\sqrt{3}} \frac{\pi}{L} \left( m -
\frac{n}{2} \right) {\bf e}_{x} + \frac{\pi}{L} n {\bf e}_{y} \;\;\;\; m,n
\in Z^{+}  &  \label{971}
\end{eqnarray}
for the differentiable functions on  the tetrahedron.

Going back to our original problem, one finds that the functions
$\Psi_{(m,n)}$ are  not by themselves solutions to equation
(\ref{95}). However, it is now an easy task to produce solutions
for the eigenvalue problem on the lattice. One has first to
split artificially the points in figure 5 in  two categories,
say black points and blank points,
depending on the orientation of the adjacent links as  shown in
the figure. Then, one may form a function $\Psi^{(1)}$ on the lattice of the
type (\ref{93}) with
\begin{eqnarray}
 &  f_{i}^{(1)} = exp( i {\bf r}_{i}{\bf \cdot k}^{(m,n)})  &
   for\;\; black \;\;points         \nonumber     \\
 &  f_{i}^{(1)} = exp(- i {\bf r}_{i}{\bf \cdot k}^{(m,n)})  &
   for\;\; blank \;\;points    \label{98}
\end{eqnarray}
Alternatively, one may also form a function $\Psi^{(2)}$ with
\begin{eqnarray}
 &  f_{i}^{(2)} = exp(- i {\bf r}_{i}{\bf \cdot k}^{(m,n)})  &
   for\;\; black\;\; points         \nonumber     \\
 &  f_{i}^{(2)} = exp( i {\bf r}_{i}{\bf \cdot k}^{(m,n)})  &
   for \;\;blank \;\;points    \label{99}
\end{eqnarray}
It is obvious that either of these two choices gives rise to
a  singlevalued function for the lattice on the tetrahedron. On
the other hand, by application of the hamiltonian on
$\Psi^{(1)}$ we get
\begin{eqnarray}
 H \Psi^{(1)} & = & \gamma \sum_{i \;\bullet} \sum_{<i,j>} exp(- i  {\bf
r}_{j}{\bf \cdot k}^{(m,n)}) a_{i}^{+} \left| O \right\rangle
 + \gamma \sum_{i \;\circ} \sum_{<i,j>} exp( i  {\bf
r}_{j}{\bf \cdot k}^{(m,n)}) a_{i}^{+} \left| O \right\rangle
  \nonumber    \\
 & = & \gamma \sum_{i \;\bullet}  exp(- i  {\bf r}_{i}{\bf \cdot k}^{(m,n)})
       \sum_{j}  exp(- i  {\bf u}_{j}{\bf \cdot k}^{(m,n)})
      a_{i}^{+} \left| O \right\rangle    \nonumber   \\
 &  &  + \: \gamma \sum_{i \;\circ}
           exp( i  {\bf r}_{i}{\bf \cdot k}^{(m,n)})
       \sum_{j}  exp( i  {\bf v}_{j}{\bf \cdot k}^{(m,n)})
      a_{i}^{+} \left| O \right\rangle \label{910}
\end{eqnarray}
where $\{ {\bf u}_{j} \}$ and $\{ {\bf v}_{j} \}$ are the two triads
mentioned in the previous section. The quantity
\begin{equation}
 \sum_{j}  exp(- i  {\bf u}_{j}{\bf \cdot k}^{(m,n)}) =
  \sum_{j}  exp( i  {\bf v}_{j}{\bf \cdot k}^{(m,n)}) =
    \lambda_{(m,n)}        \label{911}
\end{equation}
is constant over the lattice, so that $\Psi^{(1)}$ is mapped
into $\Psi^{(2)}$ by the action of the hamiltonian, and
viceversa. In this way we have built, for each pair $(m,n)$, a
two-dimensional space which is held invariant under the action
of $H$. The eigenvalues of this operator can be expressed,
therefore, in the form
\begin{equation}
E_{(m,n)} = \pm \gamma |\lambda_{(m,n)}| = \pm \gamma \sqrt{1 + 4
cos^{2}\frac{\sqrt{3}}{2} k_{x}^{(m,n)} +
4 cos \frac{\sqrt{3}}{2}k_{x}^{(m,n)} \; cos \frac{3}{2}
k_{y}^{(m,n)} }          \label{912}
\end{equation}

At this point, we may ask whether all the momenta
which satisfy the  boundary conditions in the continuum are
compatible with the quantization conditions imposed by the
lattice. These arise from the fact that, under
certain translations, the lattice maps into itself. An
independent set of them is given by the transformations $2P({\bf
T}_{1} + {\bf T}_{2})$ and $2P(- {\bf T}_{1} + 2{\bf T}_{2})$,
where $P$ is a positive integer equal to $2/3$ times the number
of hexagons along the $x$ direction.
The projections of the allowed momenta onto
these two vectors are quantized in the form
\begin{eqnarray}
 & 2P {\bf k \cdot}({\bf T}_{1} + {\bf T}_{2}) = 2 \pi p
    \;\;\;\;\; p \in Z  &   \label{913}  \\
 & 2P {\bf k \cdot}(-{\bf T}_{1} + 2{\bf T}_{2}) = 2 \pi q
    \;\;\;\;\; q \in Z  &   \label{914}
\end{eqnarray}
The momenta (\ref{971}) found in the continuum satisfy
automatically these conditions. This is easily seen after
adjusting properly the length $L$ of the sides of the triangular
faces to its lattice dimension, $3P$. We end
up with the outcome that the allowed momenta on the continuum
fill up the first Brillouin zone of the lattice. In
general, every two opposite momenta give rise to two solutions
with opposite energy on the lattice. There are only a few
exceptions to this rule, corresponding to those momenta sitting
on the boundary of the first Brillouin zone, in which the
lattice actually identifies modes corresponding to different
momenta. Only in such cases there is one mode for each pair of
opposite points in the reciprocal lattice. This explains why in
all the honeycomb lattices inscribed in the tetrahedron the
number of zero modes is two, one for each of the two independent
Fermi points. We give in table 1 the spectrum of the
independent modes contained in the first Brillouin zone, in the
particular case $P = 2$ which corresponds to the lattice of
figure 5 (we set $\gamma = -1$).
The energy eigenvalues coincide precisely with those
obtained by numerical diagonalization of the lattice
hamiltonian, represented in the top diagram of figure 7.

\begin{table}
\vskip-1.0cm
\centering
\begin{tabular}{||l|l|c||}  \hline
            $k_{x}$      & $k_{y}$        &   $E$  \\  \hline
 $ 0$                     &       0         &  $  -3 $  \\ \hline
 $ \pi /(3\sqrt{3})$  &                &  $\pm (1 + \sqrt{3})$  \\
 \cline{1-2}
  $ \pi /(6\sqrt{3})$  & $ \pi/6  $   &    \\
 \cline{1-2}
$-\pi/(6\sqrt{3})  $  &  $ \pi/6$   &     \\  \hline
  $\pi/(2\sqrt{3})   $&  $\pi/6  $  &  $  \pm \sqrt{5}$  \\
  \cline{1-2}
                      &$\pi/3$      &      \\
  \cline{1-2}
 $-\pi/(2\sqrt{3}) $   & $\pi/6$     &      \\ \hline
 $2\pi/(3\sqrt{3})$   &             & $ \pm2$  \\
 \cline{1-2}
 $\pi/(3\sqrt{3})$    & $\pi/3$     &      \\
 \cline{1-2}
$-\pi/(3\sqrt{3})$    & $\pi/3$     &      \\ \hline
 $5\pi/(6\sqrt{3})$    & $\pi/6$     &  $ \pm \sqrt{2}$   \\
\cline{1-2}
 $2\pi/(3\sqrt{3})$    & $\pi/3$     &      \\
\cline{1-2}
 $\pi/(6\sqrt{3})$    & $\pi/2$     &      \\
\cline{1-2}
$-5\pi/(6\sqrt{3})$    & $\pi/6$     &     \\
\cline{1-2}
$-2\pi/(3\sqrt{3})$    & $\pi/3$     &      \\
\cline{1-2}
$-\pi/(6\sqrt{3})$    & $\pi/2$     &      \\  \hline
$\pi/\sqrt{3}  $      &             &   $\pm 1 $  \\
\cline{1-2}
$\pi/(2\sqrt{3})$    & $\pi/2$     &     \\
\cline{1-2}
$-\pi/(2\sqrt{3})$    & $\pi/2$     &     \\ \hline
$\pi/\sqrt{3}$    & $\pi/3$     & $ 1$    \\
\cline{1-2}
                   &$ 2\pi/3$     &       \\
\cline{1-2}
$-\pi/\sqrt{3}$    & $\pi/3 $     &     \\  \hline
$7\pi/(6\sqrt{3})$    & $\pi/6$     &  $\pm(\sqrt{3} - 1)$    \\
\cline{1-2}
$\pi/(3\sqrt{3})$    & $2\pi/3$     &     \\
\cline{1-2}
$-7\pi/(6\sqrt{3})$    & $\pi/6$     &     \\ \hline
$4\pi/(3\sqrt{3})$     &             &  $0$   \\
\cline{1-2}
$2\pi/(3\sqrt{3})$     & $2\pi/3$    &      \\ \hline
\end{tabular}
\vskip-1.0cm
\end{table}

The method we have just illustrated in the case of the
tetrahedron can be  applied to solve the spectrum of honeycomb
lattices inscribed in any of the remaining orbifolds, namely the
octahedron and the icosahedron. The lesson that we learn by
following this approach is that the eigenfunctions of the curved
lattice are given by momenta still lying in the first Brillouin
zone, in which some points have been identified according to the
symmetries of the lattice  unfolded on the plane. In the case of
 the icosahedron, however, one may devise
more economic techniques in order to predict the relevant
properties for solid state applications. It is  possible, as we
will see in what follows, to extract the essentials of the
method described above, in order to model the properties of the
modes near the Fermi points to a high degree of approximation.

\section{Effective field theories in the continuum limit}

In this section, rather than pursuing the exact diagonalization
of the hamiltonian (\ref{91}) we undertake the formulation of
effective field theories describing the low energy excitations
of the folded honeycomb lattices at half-filling.
The analysis of the previous section
indicates the general way to proceed, in a manner which makes
possible the treatment of the more complicated lattice on the
icosahedron. The resolution of the honeycomb lattice on the
tetrahedron shows that the formal expression of the dispersion
relation remains untouched in the curved lattice, and
susceptible of being considered in the continuum limit to
produce a simple field theory. Taking the continuum limit is a
way of amplifying the structure of the levels infinitely close
to each of the Fermi points. For this reason, the states of the
effective  field theory are attached to any of the two
independent Fermi points, giving rise in the case of the planar
lattice to the spectrum  of two uncoupled Dirac spinors.
Regarding folded lattices in which the two sublattices of black
and blank points are exchanged by going around a conical
singularity, the admissible wavefunctions are made of pairs of  plane
waves with opposite momenta. We have already applied this
construction for the lattice on the tetrahedron, and it turns
out to be also pertinent for the lattice on the icosahedron.
In momentum space the inversion with respect to the origin
exchanges the two independent classes of Fermi points. It
becomes clear that, for the mentioned honeycomb lattices, the
states of the theory have to accomodate into the spectrum of two
{\em coupled} Dirac spinors.

To understand the nature of the interaction between the two
spinors we may have a deeper look at the process of
diagonalization of the lattice hamiltonian. The boundary
conditions imposed in (\ref{961})-(\ref{964}) are a resort to
obviate the fact that, in principle, two charts are needed to cover
each of the conical singularities. Focusing on the
singularity at the origin, for instance, two appropriate local
coordinate systems are depicted in figure 6. In each separate
coordinate patch the two wavefunctions
\begin{eqnarray}
\Psi_{\bullet}   & = &  \sum_{i \;\bullet}
\mbox{\Large $e^{i{\bf k \cdot r}_{i}}$} a_{i}^{+}
\left| O \right\rangle        \\
\Psi_{\circ}     & = &  \sum_{i \;\circ}
\mbox{\Large $e^{i{\bf k \cdot r}_{i}}$}a_{i}^{+}
\left| O \right\rangle
\end{eqnarray}
span a two dimensional invariant subspace in which the effective
hamiltonian is the same as in (\ref{9})
\begin{equation}
{\cal H}_{+} =     \left(
\begin{array}{cc}
 0  &  \gamma \sum_{j}\mbox{\Large $e^{i{\bf k \cdot u}_{j}}$} \\
\gamma \sum_{j}\mbox{\Large $e^{i{\bf k \cdot v}_{j}}$} &  0
\end{array} \right) \label{81}
\end{equation}
A similar pair of wavefunctions with the opposite momentum gives
rise to a second effective hamiltonian
\begin{equation}
{\cal H}_{-} =     \left(
\begin{array}{cc}
 0  & \gamma \sum_{j}\mbox{\Large $e^{-i{\bf k \cdot u}_{j}}$} \\
 \gamma \sum_{j}\mbox{\Large $e^{-i{\bf k \cdot v}_{j}}$} &  0
\end{array} \right) \label{82}
\end{equation}
The respective regions to the left of the two coordinate patches
are in correspondence by an appropriate map, and the important
point is that under this mapping every vector in the tangent
space suffers a rotation of $\pi$. This applies in particular to
the momenta, so that when going from the left of region I to the
left of region II in figure 6 the two effective hamiltonians
(\ref{81}) and (\ref{82}) are exchanged.

In the continuum limit, the rotation of the momenta has the
following consequences. A momentum ${\bf k}$ about the Fermi
level like (\ref{66}) is mapped into
\begin{equation}
-{\bf k} = -\frac{4\pi }{3\sqrt{3}} {\bf e}_{x} - \mbox{\boldmath $\delta k$}
\end{equation}
This implies that an operator like ${\cal H}_{+}$ in the
continuum limit
\begin{equation}
\lim_{a \rightarrow 0} {\cal H}_{+}/a =  \left.
- \frac{3}{2}\gamma \: \mbox{\boldmath
$\sigma$}^{T}\mbox{\boldmath $\cdot \delta k$}
\right|_{{\bf k} = (4\pi /3\sqrt{3}) {\bf e}_{x}} \label{71}
\end{equation}
goes by the mentioned mapping into the operator
\begin{equation}
\lim_{a \rightarrow 0} {\cal H}_{-}/a =  \left.
 \frac{3}{2} \gamma \:\mbox{\boldmath
$\sigma$}^{T}\mbox{\boldmath $\cdot \delta k$}
\right|_{{\bf k} = -(4\pi /3\sqrt{3}) {\bf e}_{x}} \label{72}
\end{equation}
The rotation suffered by \mbox{\boldmath $\delta k$} is of the
kind
produced by the spin connection for the curvature accumulated at
the conical singularity. However, the change of Fermi point when going from
(\ref{71}) to (\ref{72}) has a different character, and has to
be dealt with separately  by means of a different connection.

The mapping between the two local coordinate systems I and II
exchanges excitations which are bounded respectively to two
independent Fermi points. In the continuum limit this means the
exchange of the two Dirac spinors, and this kind of twist can be
realized by a proper connection in the internal space of these
two fields. In general, the connection can be given support on a set of
disjoint segments linking pairs of neighboring singularities on
the two-dimensional surface. On the tetrahedron and the
icosahedron we may form, respectively, two and six of such
cuts, across of which the rotation in the internal space of the
two
spinors takes place. More physically, the rotation can be
implemented by inserting a line of magnetic flux at each of the
conical singularities. The flux has to be nonabelian and
properly adjusted to produce the twist by going around the
puncture. A correct parametrization is achieved by introducing
the $SU(2)$-valued connection
\begin{equation}
{\bf A} \equiv {\bf A}^{(a)} \tau^{(a)} \;\;\;\;\;\;\;a = 1, 2, 3
\end{equation}
with $\{ \tau^{(a)} \}$ being the three Pauli matrices.
The only nonvanishing component may be taken to be, in local polar
coordinates around each puncture,
\begin{equation}
A_{\phi} = \frac{\Phi }{2\pi } \tau^{(2)}
\end{equation}
By setting the magnetic flux to $\Phi = \frac{\pi}{2}$, the
nonabelian phase acquired by the doublet of spinors after going
around each conical singularity is, as required,
\begin{equation}
\mbox{\Large $e^{i\oint {\bf A}}$} =  \left(
\begin{array}{cc}
 0  &  1 \\
 -1 &  0
\end{array} \right)
\end{equation}
This picture implies the existence of a fictitious magnetic
monopole inside the surface. Its charge $g$ can be computed by
adding up the individual fluxes of all the lines
\begin{equation}
g = \frac{1}{4\pi } \sum_{i = 1}^{N} \frac{\pi }{2} = \frac{1}{8}N
\end{equation}
$N$ being the number of conical singularities on the surface. It
is worth mentioning that the values of $g$ required for the
tetrahedron and the icosahedron are 1/2 and 3/2, respectively,
and therefore compatible with the standard quantization
condition of the monopole charge\cite{col}.

To summarize, we have developed a picture in which the continuum
limit for honeycomb lattices
on the tetrahedron and the icosahedron at half-filling is given
by the effective field theory of
a doublet of spinors interacting with the curvature and color
magnetic fields accumulated at a certain number of conical
singularities. Although the exact resolution of this model is
beyond the scope of the present paper, it is possible to show with much
less effort that it reproduces the low energy spectra and
correct numbers of zero modes for the mentioned lattices in the
limit of large number of points. As a first approximation, one
may consider the model in which both curvature and magnetic
field are made uniform over the two-dimensional surface. Then
the system becomes that of a couple of spinors on the sphere
with magnetic monopole field. The spectrum is
obtained by solving the eigenvalue problem for the covariant
Dirac operator\cite{bd}
\begin{equation}
i \sigma^{a} e^{\mu}_{a}\left(\nabla_{\mu} - iA_{\mu}\right) \Psi_{n} =
\varepsilon_{n} \Psi_{n} \;\;\;\;\; a,\mu = 1, 2
\end{equation}
where $e^{\mu}_{a}$ is the zweibein for the sphere and, in
spherical coordinates,
\begin{eqnarray}
\nabla_{\theta}  & = & \partial_{\theta}   \\
\nabla_{\phi}    & = & \partial_{\phi} - \frac{1}{4}\:[\sigma^{1},
     \sigma^{2}] \:cos \theta               \\
A_{\theta}       & = &  0                 \\
A_{\phi}         & = &  g \: cos \theta \: \tau^{(2)}
\end{eqnarray}
The Dirac operator can be diagonalized by introducing the
angular momentum of the whole system made of spinor fields,
magnetic field and curvature. The total angular momentum
operators turn out to be, for the lower spinor component
$\Psi_{\downarrow}$,
\begin{eqnarray}
J_{\pm}    & = & \pm \: \mbox{\Large $e^{\pm i \phi}$} \: \nabla_{\theta}
    + i \: \mbox{\Large $e^{\pm i \phi}$} \: \frac{cos \theta}{sin \theta}
\:  \left(\nabla_{\phi} - ig \: cos \theta \: \tau^{(2)}\right)
  -  \mbox{\Large $e^{\pm i\phi}$} \: sin\theta
      \:   \left(\frac{1}{2} - g \tau^{(2)}\right)         \\
J_{z}      & = & -i\left(\nabla_{\phi} - ig \: cos \theta \: \tau^{(2)}\right)
     - \: cos \theta \: \left(\frac{1}{2} - g \tau^{(2)}\right)
\end{eqnarray}
For the upper component $\Psi_{\uparrow}$, the corresponding operators
are similar except for a change of sign in front of the 1/2 fractions.
By squaring
the Dirac operator, each of the spinor components comes to obey
the equation, with respective angular momentum operators,
\begin{equation}
\left({\bf J}^{2} + \frac{1}{4} - g^{2}\right) \Psi_{n} = \varepsilon_{n}^{2}
r^{2} \Psi_{n}
\end{equation}
where $r$ parametrizes the radius of the sphere. The spectrum is
given in terms of the angular momentum quantum number $j$
\begin{equation}
\varepsilon_{j}^{2} r^{2} = j\left(j + 1\right) + \frac{1}{4} - g^{2} =
     \left(j + \frac{1}{2}\right)^{2} - g^{2}    \label{51}
\end{equation}
As is well known, there is a minimum value of the angular
momentum $j$ dictated by $g$, so that the number of zero modes in the
spectrum depends exclusively on the value of the monopole charge.

Let us specialize now to the models corresponding respectively
to the tetrahedron and the icosahedron. In the first case the
monopole charge required is $g = 1/2$, and the model predicts
the existence of two zero modes, one for each spinor component,
in the continuum limit of the honeycomb lattice at half-filling.
This result can be tested with the spectra obtained by numerical
diagonalization of the hamiltonian (\ref{91}). Let us call generically a
honeycomb lattice inscribed on the tetrahedron by $\Theta_{m}$,
$m$ being the total number of points in the lattice. These
lattices can be ordered along a sequence of increasing $m$,
whose general term is of the form $\Theta_{12(n + 1)^{2}}\;,\;
n \in Z^{+}$. We have depicted the result of the numerical
diagonalization of three of these lattices
in figure 7 (we take units in which $\gamma = -1$). In all
cases, two states appear with an energy compatible with zero
within the computer precision, showing that our mean field
approximation already produces the correct number of zero modes.
Furthermore, from
$\Theta_{192}$ on the next low energy states are given by two
close triplets above the Fermi level and two other with the
opposite energy. This bears a reasonable agreement with what
is expected from the formula (\ref{51}). The highest
dimension of an irreducible representation for the tetrahedron
symmetry group is three and, consequently, a significant departure from
spherical symmetry develops above $j = 1$ in the spectrum.

We deal similarly with the honeycomb lattices folded on the
icosahedron. The effective field theory demands now a monopole
charge $g = 3/2$. Relying again on the spherical approximation
to compute the number of zero modes, we come out
with the theoretical prediction that there should be a couple of
triplets lying at zero energy, in the continuum limit.
We denote the fullerene lattices by $C_{m}$ according to the
total number of lattice points $m$. The sequence with increasing
number of points is given by the general term $C_{60(n + 1)^{2}}\;,\;
n \in Z^{+}$. We have represented in figure 8 the spectra
obtained by numerical diagonalization of three such lattices (we
have set again $\gamma = -1$).
Although we do not find as before any zero modes from the start,
there is clear evidence that the couple of triplets close to zero
energy approach asymptotically the origin of the energy
scale in the limit of large lattices. The levels given by the
formula (\ref{51}) are in good agreement with those in the numerical
spectra up to the point in which the highest dimension of an
irreducible representation of the icosahedron symmetry group is
reached. This happens for $j = 2$. The levels next to the couple
of quintuplets at each  side of the spectrum are a couple of
quadruplets and other of triplets, which may be thought as
arising from the split of two $j = 3$ multiplets by breaking
down to the icosahedron symmetry group.

The results of the numerical diagonalization of the lattice
hamiltonian (\ref{91}) support the correctness of the effective
field theory developed to account for the continuum limit of the
curved lattices. Let us mention, finally, that there is a
similar sequence of honeycomb lattices folded on the octahedron.
In all of them one may define consistently two sublattices of
black and blank points, respectively, over the whole surface.
According to our picture, the continuum limit for these lattices
at half-filling should be given by a field theory on the
octahedron in the absence of magnetic field. The inspection of
the numerical spectra for these lattices shows that none of them
has any mode sitting at zero energy, nor this value is
approached asymptotically in the continuum limit. It is
reassuring to find that this is, in fact, the prediction
obtained in the framework of the theoretical model after
switching off the magnetic field.

\section{Summary and prospects}

The leitmotif of this paper was to explain the details of how
a continuum model can be used to study the electronic and elastic
properties of the fullerene molecules \cite{nos}. Along the way we
have found some interesting points that we will highlight and
comment here.

This work is based on two fundamental points. The
first one is the existence of isolated Fermi points
instead of lines in the graphite molecule at half filling.
This leads to the formulation of a continuum limit for describing
the spectrum of electronic excitations around any of these points.
We have seen that the particular geometry
of the lattice determines the continuum model to
be that of two massless, independent, free, Dirac spinors.

The second basic fact in this paper is the existence of bigger
fullerene molecules
derived from a given geometry and the regularity found
in their spectrum of excitations around the Fermi level. That is what
allows us to propose the validity for them
of the continuum limit found in
the -- infinite -- graphite lattice.

Once we know the properties of the flat graphite lattice,
the next interesting point refers to the boundary conditions that
can be imposed on the lattice without destroying its main features.
At this respect it is worth noticing that the standard procedure
to study the properties of Bravais
latices assumes most of the times the choice of Von Karman , i.e.,
periodic, boundary conditions. While it is obvious that any choice
of {\it flat} boundary conditions should not alter the main properties
of the bulk lattice, we must be careful when considering boundary
conditions of the type described in section 2. Despite the innocent
presentation of figure 5, the identifications done to
form the polihedra are curving the lattice -- as it is clear
from figure 6 -- and this is a non trivial operation. The simplest
way to look at it is to think on the way that curvature is induced
on the exagonal tesselation. This is done by substituting some exagons by
n-gons with n less than six. Such substitution will, in general,
induce frustration on the Bravais lattice that underlies the
exagonal lattice. In particular, for odd n, the two triangular
sublattices are interchanged. It is then not trivial
that we could successfully
complete the study of the tetrahedron.

Before leaving the subject of the
boundary conditions, let us mention some results
concerning the simplest
situation, namely, the case of a graphite sheet with periodic
boundary conditions. When the lattice is folded so as to
form a torus, i.e., a compact surface with no intrinsic curvature,
one finds that the electronic spectrum is precisely the one predicted
by the continuum model: the doubled spectrum of a Dirac spinor in a
plane with doubly periodical boundary conditions. What we observe
numerically is a structure of cuadruplets (one doublet for each
Dirac spinor) equally spaced in energy. We end this comment by
mentioning that cylindrical shapes and tubular structures closed
at the extremes with conical singularities (capsules), have been
described in \cite{cap}.

The phenomenological model described in section 4 summarizes all
the features discussed so far. We are led to study two Dirac
spinors on compact surfaces with curvature singularities and
edges of frustration.
The sphere takes into account the compactness and
curvature of the lattice, and the monopole mimics the frustration.

We envisage various directions in which this work can be continued.
The first concerns the Statistical Mechanics applications.
We have seen how, starting from the fullerene molecules,
we came to
the study of an entirely new family of two dimensional lattices where
one can solve the spectrum of the hopping hamiltonian: honeycomb lattices
folded and wrapped around truncated regular polyhedra. Any exactly solvable
model in Statistical Mechanics has some interest on its own, regardless
of its immediate applicability to physical problems. In our case, we came
the other way around as we took our models directly from existing physical
examples. Let us notice that although
some of this lattices have already appeared in the literature \cite{sam,tetr},
they were formed by a fixed number of points. The main novelty here lies in
the fact that our lattices can grow while preserving the coordination and
the global symmetry group so that it makes sense to study the continuum and
thermodynamical limits. {\it We then have two dimensional Statistical models
defined on curved, compact surfaces}.
Work on the study of the thermodynamic
limit of two dimensional
models such as the Ising or Hubbard models defined in
the new lattices is currently in progress.

As for the Solid State implications,
the universal character of the method proposed allows its
immediate applicability to the study of the electronic spectrum
of all kind of fullerenes existing (as the family of the
elliptical $C_{70}$) or proposed. Among those,
the geometries with negative curvature \cite{neg}
may present new challenges as they live in
curved but non compact surfaces.
The goemetries proposed so far, however, have the adventage
of building a three dimensional lattice, so that the
problem needs only to be solved within one unit cell.
The existence of odd numbered rings, which now are heptagons,
exchanges the two sublattices of the graphite structure.
Thus, following the analysis of this paper, a fictitous
gauge field needs to be introduced.
The detailed nature of this field is postponed to a future
publication. On the other hand, the fact that the
square of the Dirac equation is the Laplace equation,
and the use of periodic boundary conditions, allows
us to make some general remarks about the spectrum
of these systems\cite{nos}.

Finally, we leave out of this paper a discussion of the
{\it elastic} modes of these molecules. For simple central
force models, the projection of the structure on a plane,
and the imposition of non trivial boundary conditions
can be generalized in a straightforward way. The main difference
is that the low wavelength accoustical modes are related
to the center of the Brillouin zone of the flat graphite sheets.
For these excitations, frustration does not play the same
striking role as at the corners of the Brillouin zone.
Thus, we expect that these modes will be well described by
the standard theory of elasticity of curved shells.

\vspace{1cm}
{\bf Acknowledgements.} It is a pleasure to thank G. Sierra for
interesting discusions on the subject of the hopping hamiltonian.
M.A.H.V. thanks the department of F\'{\i}sica Te\'orica of the
Universidad Aut\'onoma of Madrid for finantial support during the
course of this work.
This work has been partially supported by the CICyT (Spain).

\newpage
\begin{figure}[p]
\centering
\setlength{\unitlength}{1.2cm}
\begin{picture}(15,13)(2,0)

\multiput(5.66,2)(0,3){3}{\line(5,-3){0.83}}
\multiput(6.5,1.5)(0,3){3}{\line(5,3){0.83}}
\multiput(7.33,5)(0,3){2}{\line(5,-3){0.83}}
\put(8.16,7.5){\line(5,3){0.83}}
\put(9,5){\line(5,-3){0.83}}
\put(9.83,4.5){\line(5,3){0.83}}
\put(10.66,5){\line(5,-3){0.83}}
\put(11.5,4.5){\line(5,3){0.83}}

\multiput(5.66,3)(0,3){3}{\line(5,3){0.83}}
\multiput(6.5,3.5)(0,3){3}{\line(5,-3){0.83}}
\multiput(7.33,3)(0,3){2}{\line(5,3){0.83}}
\put(8.16,3.5){\line(5,-3){0.83}}

\put(9,6){\line(5,3){0.83}}
\put(9.83,6.5){\line(5,-3){0.83}}
\put(10.66,6){\line(5,3){0.83}}
\put(11.5,6.5){\line(5,-3){0.83}}

\multiput(6.5,3.5)(0,3){2}{\line(0,1){1}}
\multiput(8.16,3.5)(0,3){2}{\line(0,1){1}}
\multiput(9.83,3.5)(0,3){2}{\line(0,1){1}}

\multiput(5.66,2)(0,6){2}{\line(0,1){1}}
\multiput(7.33,2)(0,6){2}{\line(0,1){1}}
\put(9,5){\line(0,1){1}}
\put(10.66,5){\line(0,1){1}}
\put(12.33,5){\line(0,1){1}}
\multiput(5.66,1)(0,1){9}{\line(0,1){0.4}}
\multiput(5.66,1.6)(0,1){9}{\line(0,1){0.4}}

\multiput(13.16,5.5)(-0.833,-0.5){9}{\line(-5,-3){0.333}}
\multiput(12.66,5.2)(-0.833,-0.5){9}{\line(-5,-3){0.333}}
\multiput(13.16,5.5)(-0.833,0.5){9}{\line(-5,3){0.333}}
\multiput(12.66,5.8)(-0.833,0.5){9}{\line(-5,3){0.333}}
\put(7.33,5){\line(0,1){0.4}}
\put(7.33,6){\line(0,-1){0.4}}
\put(8.16,4.5){\line(5,3){0.333}}
\put(9,5){\line(-5,-3){0.333}}
\put(8.16,6.5){\line(5,-3){0.333}}
\put(9,6){\line(-5,3){0.333}}
%0

\end{picture}
\caption{Generic triangular block for honeycomb lattices folded on the
tetrahedron, the octaedron and the icosahedron.}
\end{figure}
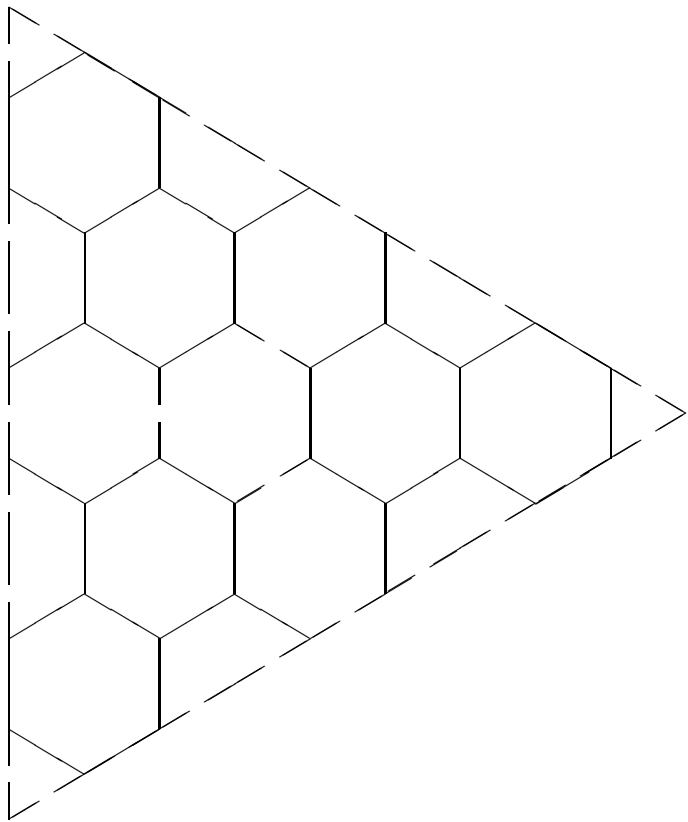

\newpage
\begin{figure}[p]
\centering
\setlength{\unitlength}{1.2cm}
\begin{picture}(15,13)(2,0)

\multiput(4.83,1.5)(0,3){3}{\line(5,3){0.83}}
\multiput(5.66,2)(0,3){3}{\line(5,-3){0.83}}
\multiput(6.5,1.5)(0,3){3}{\line(5,3){0.83}}
\multiput(7.33,2)(0,3){3}{\line(5,-3){0.83}}
\multiput(8.16,1.5)(0,3){3}{\line(5,3){0.83}}
\multiput(9,2)(0,3){3}{\line(5,-3){0.83}}
\multiput(9.83,1.5)(0,3){3}{\line(5,3){0.83}}
\multiput(10.66,2)(0,3){3}{\line(5,-3){0.83}}
\multiput(11.5,1.5)(0,3){3}{\line(5,3){0.83}}
\multiput(12.33,2)(0,3){3}{\line(5,-3){0.83}}

\multiput(4.83,3.5)(0,3){3}{\line(5,-3){0.83}}
\multiput(5.66,3)(0,3){3}{\line(5,3){0.83}}
\multiput(6.5,3.5)(0,3){3}{\line(5,-3){0.83}}
\multiput(7.33,3)(0,3){3}{\line(5,3){0.83}}
\multiput(8.16,3.5)(0,3){3}{\line(5,-3){0.83}}

\multiput(9,3)(0,3){3}{\line(5,3){0.83}}
\multiput(9.83,3.5)(0,3){3}{\line(5,-3){0.83}}
\multiput(10.66,3)(0,3){3}{\line(5,3){0.83}}
\multiput(11.5,3.5)(0,3){3}{\line(5,-3){0.83}}
\multiput(12.33,3)(0,3){3}{\line(5,3){0.83}}
\multiput(4.83,3.5)(0,3){2}{\line(0,1){1}}
\multiput(6.5,3.5)(0,3){2}{\line(0,1){1}}
\multiput(8.16,3.5)(0,3){2}{\line(0,1){1}}
\multiput(9.83,3.5)(0,3){2}{\line(0,1){1}}
\multiput(11.5,3.5)(0,3){2}{\line(0,1){1}}
\multiput(13.16,3.5)(0,3){2}{\line(0,1){1}}

\multiput(5.66,2)(0,3){3}{\line(0,1){1}}
\multiput(7.33,2)(0,3){3}{\line(0,1){1}}

\multiput(10.66,2)(0,3){3}{\line(0,1){1}}
\multiput(12.33,2)(0,3){3}{\line(0,1){1}}
\multiput(5.66,2)(0,3){3}{\circle*{0.2}}
\multiput(7.33,2)(0,3){3}{\circle*{0.2}}
\multiput(9,2)(0,3){3}{\circle*{0.2}}

\multiput(10.66,2)(0,3){3}{\circle*{0.2}}
\multiput(12.33,2)(0,3){3}{\circle*{0.2}}

\multiput(5.66,3)(0,3){3}{\circle{0.2}}
\multiput(7.33,3)(0,3){3}{\circle{0.2}}
\multiput(9,3)(0,3){3}{\circle{0.2}}

\multiput(10.66,3)(0,3){3}{\circle{0.2}}
\multiput(12.33,3)(0,3){3}{\circle{0.2}}

\multiput(4.83,1.5)(0,3){3}{\circle{0.2}}
\multiput(6.5,1.5)(0,3){3}{\circle{0.2}}
\multiput(8.16,1.5)(0,3){3}{\circle{0.2}}

\multiput(9.83,1.5)(0,3){3}{\circle{0.2}}
\multiput(11.5,1.5)(0,3){3}{\circle{0.2}}
\multiput(13.16,1.5)(0,3){3}{\circle{0.2}}

\multiput(4.83,3.5)(0,3){3}{\circle*{0.2}}
\multiput(6.5,3.5)(0,3){3}{\circle*{0.2}}
\multiput(8.16,3.5)(0,3){3}{\circle*{0.2}}

\multiput(9.83,3.5)(0,3){3}{\circle*{0.2}}
\multiput(11.5,3.5)(0,3){3}{\circle*{0.2}}
\multiput(13.16,3.5)(0,3){3}{\circle*{0.2}}
\thicklines

\put(9,0){\line(0,1){11}}
\put(3,5){\line(1,0){12}}
\put(8.5,10.5){$y$}
\put(14.5,4.6){$x$}

\put(6.5,6.5){\line(0,1){1}}
\put(5.66,6){\line(5,3){0.83}}
\put(6.5,6.5){\line(5,-3){0.83}}
\put(7.33,3){\line(0,-1){1}}
\put(6.5,3.5){\line(5,-3){0.83}}
\put(7.33,3){\line(5,3){0.83}}
\put(10.66,6.7){${\bf T}_{1}$}
\put(12.23,7.1){${\bf T}_{2}$}
\put(6.7,7){${\bf u}_{1}$}
\put(5.7,6.4){${\bf u}_{2}$}
\put(6.8,5.9){${\bf u}_{3}$}
\put(7.53,2.3){${\bf v}_{1}$}
\put(6.6,3.05){${\bf v}_{3}$}
\put(7.55,3.45){${\bf v}_{2}$}

%1

\end{picture}
\caption{The planar honeycomb lattice.}
\end{figure}
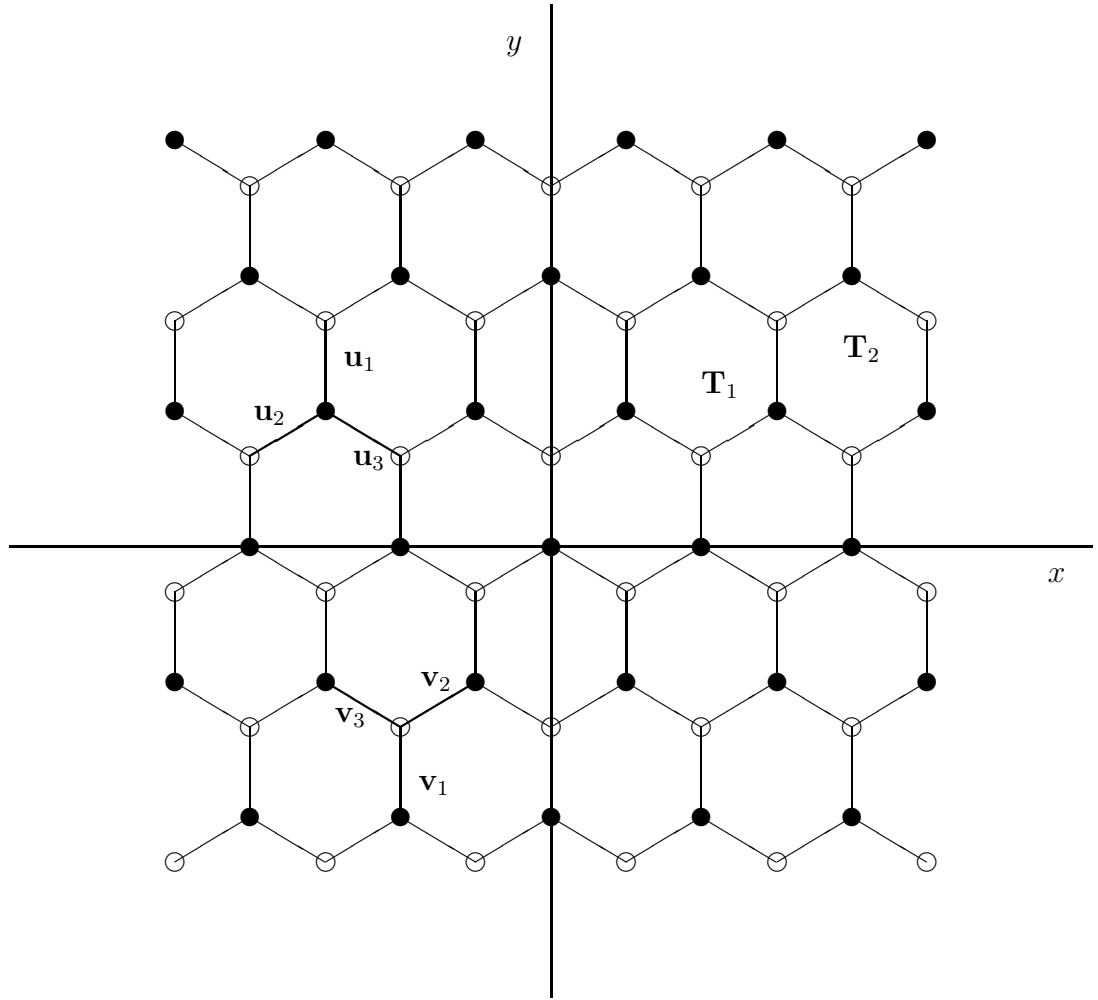

\newpage
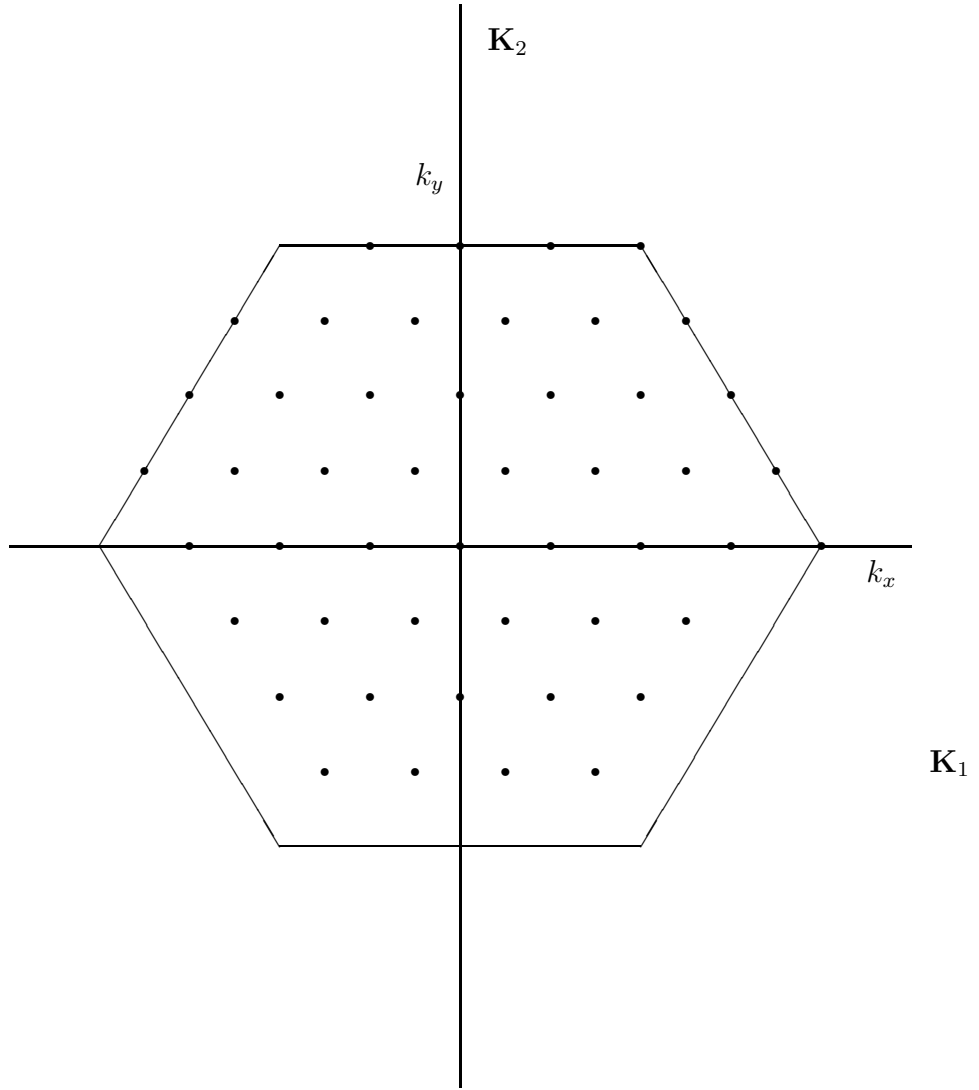
\begin{figure}[p]
\centering
\setlength{\unitlength}{1.2cm}
\begin{picture}(15,13)(2,0)
\thicklines
\put(4,6){\line(1,0){10}}
\put(9,0){\line(0,1){12}}
\put(8.5,10){$k_{y}$}
\put(13.5,5.6){$k_{x}$}
\thinlines
\put(13,6){\line(-3,5){2}}
\put(13,6){\line(-3,-5){2}}
\put(5,6){\line(3,5){2}}
\put(5,6){\line(3,-5){2}}
\put(7,9.33){\line(1,0){4}}
\put(7,2.67){\line(1,0){4}}
\multiput(8,9.33)(1,0){4}{\circle*{0.1}}
\multiput(6.5,8.5)(1,0){6}{\circle*{0.1}}
\multiput(6,7.67)(1,0){7}{\circle*{0.1}}
\multiput(5.5,6.83)(1,0){8}{\circle*{0.1}}
\multiput(6,6)(1,0){8}{\circle*{0.1}}
\multiput(6.5,5.17)(1,0){6}{\circle*{0.1}}
\multiput(7,4.33)(1,0){5}{\circle*{0.1}}
\multiput(7.5,3.5)(1,0){4}{\circle*{0.1}}
\thicklines
\put(15,2.67){.}
\put(9,12.67){.}
\put(9.3,11.5){${\bf K}_{2}$}
\put(14.2,3.5){${\bf K}_{1}$}
%
%2
\end{picture}
\caption{The reciprocal lattice in momentum space.
The hexagon depicted in the
figure represents  the first
Brillouin zone. The dots stand for independent energy eigenstates of the
lattice in figure 4.}
\end{figure}

\newpage
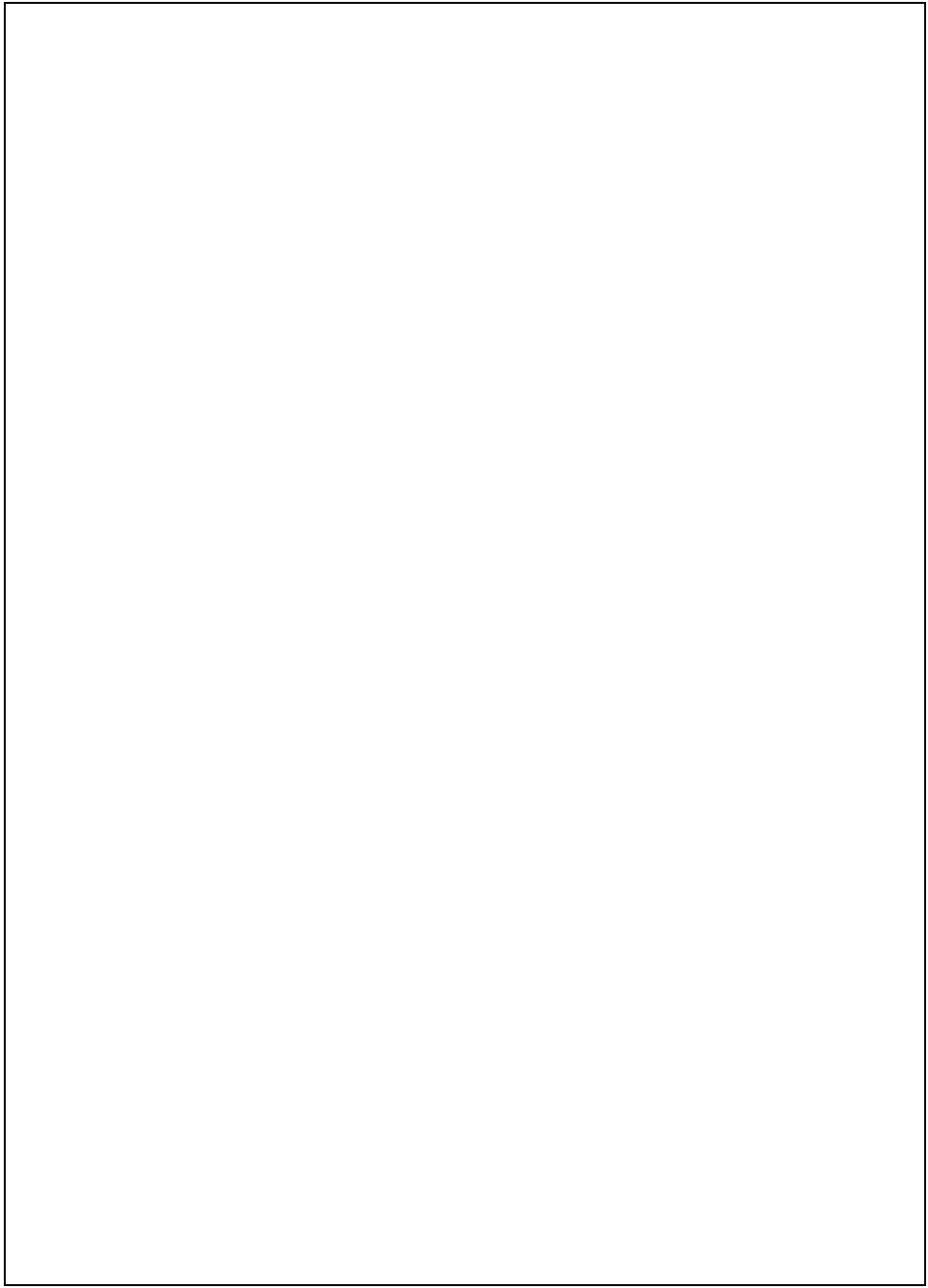
\begin{figure}[p]
\centering
%\vspace{20.0cm}
\setlength{\unitlength}{1cm}
\begin{picture}(14.2,19)
\put(0.8,1.5){\framebox(12.8,17.85)}
%
%3
\end{picture}
\caption{Representation in $(E,{\bf k})$ space of the lower
branch of the electronic dispersion relation ($\gamma
= -1$). The
cusps appear at the six corners of the first Brillouin zone.}

\end{figure}

\newpage
\begin{figure}[p]
\centering
\setlength{\unitlength}{1cm}
\begin{picture}(15,16)(-1,0)
\thicklines
\put(4,1){\line(1,0){5}}
\put(4,13){\line(1,0){5}}
\put(4,1){\line(0,1){15}}
\put(9,1){\line(0,1){12}}
\put(4,7){\line(1,0){8}}
\put(11.5,6.6){$x$}
\put(3.5,15.5){$y$}
\thinlines
\multiput(4,1)(0,6){2}{\line(5,3){5}}
\multiput(4,7)(0,6){2}{\line(5,-3){5}}
\multiput(4,2)(0,3){4}{\line(5,-3){0.83}}
\multiput(4.83,1.5)(0,3){4}{\line(5,3){0.83}}
\multiput(5.66,2)(0,3){4}{\line(5,-3){0.83}}
\multiput(6.5,1.5)(0,3){4}{\line(5,3){0.83}}
\multiput(7.33,2)(0,3){4}{\line(5,-3){0.83}}
\multiput(8.16,1.5)(0,3){4}{\line(5,3){0.83}}

\multiput(4,3)(0,3){4}{\line(5,3){0.83}}
\multiput(4.83,3.5)(0,3){4}{\line(5,-3){0.83}}
\multiput(5.66,3)(0,3){4}{\line(5,3){0.83}}
\multiput(6.5,3.5)(0,3){4}{\line(5,-3){0.83}}
\multiput(7.33,3)(0,3){4}{\line(5,3){0.83}}
\multiput(8.16,3.5)(0,3){4}{\line(5,-3){0.83}}
\multiput(4.83,3.5)(0,3){3}{\line(0,1){1}}
\multiput(6.5,3.5)(0,3){3}{\line(0,1){1}}
\multiput(8.16,3.5)(0,3){3}{\line(0,1){1}}
\multiput(5.66,2)(0,3){4}{\line(0,1){1}}
\multiput(7.33,2)(0,3){4}{\line(0,1){1}}
\multiput(4.83,1)(1.67,0){3}{\line(0,1){0.5}}
\multiput(4.83,13)(1.67,0){3}{\line(0,-1){0.5}}
\multiput(4,2)(0,3){4}{\circle*{0.2}}
\multiput(5.66,2)(0,3){4}{\circle*{0.2}}
\multiput(7.33,2)(0,3){4}{\circle*{0.2}}
\multiput(9,2)(0,3){4}{\circle*{0.2}}
\multiput(4,3)(0,3){4}{\circle{0.2}}
\multiput(5.66,3)(0,3){4}{\circle{0.2}}
\multiput(7.33,3)(0,3){4}{\circle{0.2}}
\multiput(9,3)(0,3){4}{\circle{0.2}}
\multiput(4.83,1.5)(0,3){4}{\circle{0.2}}
\multiput(6.5,1.5)(0,3){4}{\circle{0.2}}
\multiput(8.16,1.5)(0,3){4}{\circle{0.2}}
\multiput(4.83,3.5)(0,3){4}{\circle*{0.2}}
\multiput(6.5,3.5)(0,3){4}{\circle*{0.2}}
\multiput(8.16,3.5)(0,3){4}{\circle*{0.2}}
\put(4,7){\oval(1.5,1.5)[l]}
\put(9,4){\oval(1.5,1.5)[r]}
\put(9,10){\oval(1.5,1.5)[r]}
\put(9,13){\oval(2,2)[t]}
\put(9,1){\oval(2,2)[b]}
\put(10,1){\line(0,1){12}}
%
%4

\end{picture}
\caption{Curved honeycomb lattice unfolded on the plane.
The outer lines indicate the identifications between boundary
segments which embed the lattice on the tetrahedron.}
\end{figure}
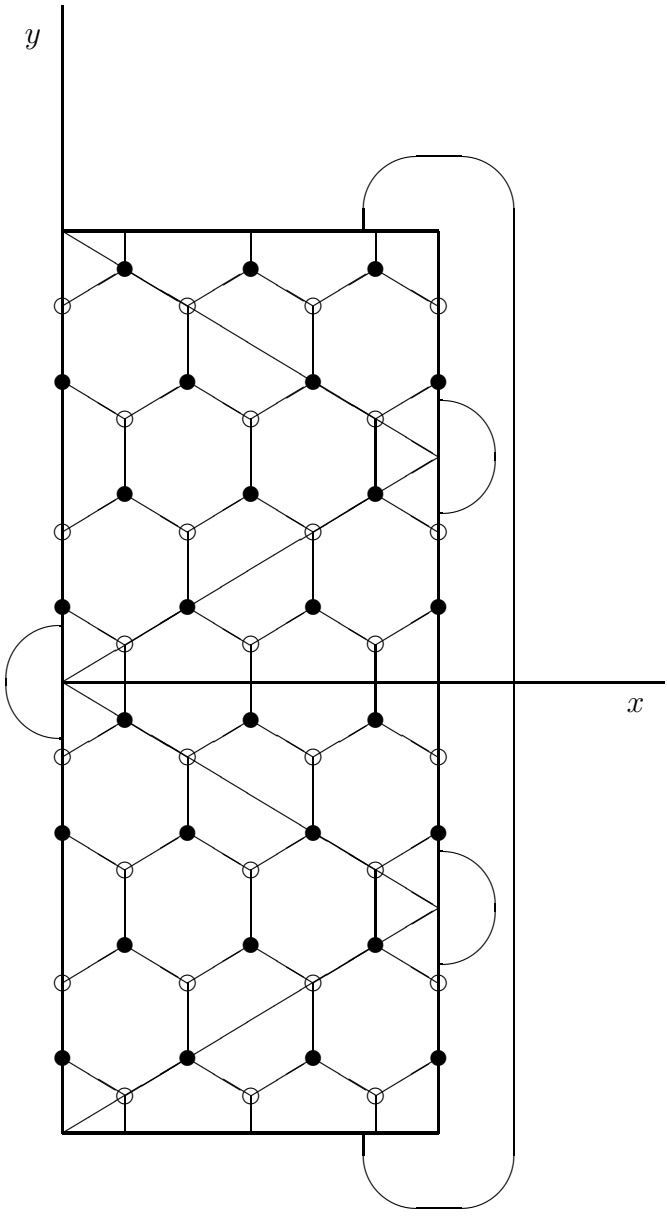

\newpage
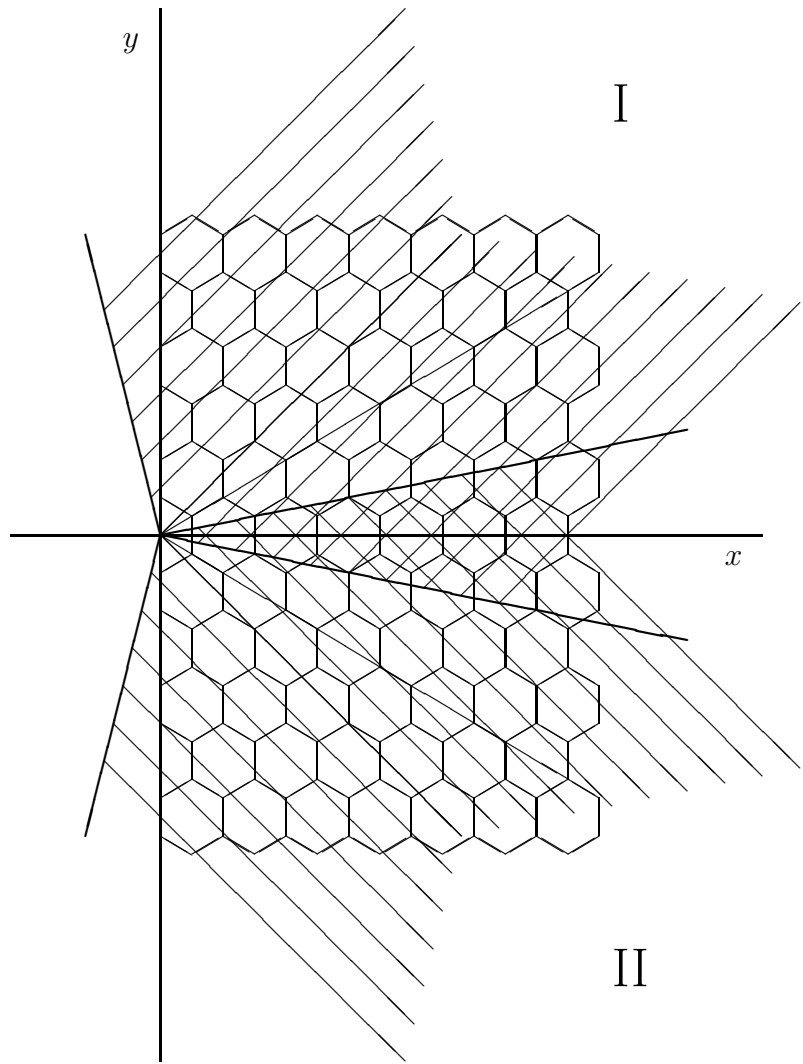
\begin{figure}[p]
\centering
\setlength{\unitlength}{1cm}
\begin{picture}(16,16)(-1,0)
\thicklines
\put(4,0){\line(0,1){14}}

\put(2,7){\line(1,0){10}}
\put(11.5,6.6){$x$}
\put(3.5,13.5){$y$}
\put(10,12.5){\LARGE I}
\put(10,1){\LARGE II}

\thinlines

\multiput(4,3)(0,1.5){6}{\line(5,-3){0.417}}
\multiput(4.417,2.75)(0,1.5){6}{\line(5,3){0.417}}
\multiput(4.83,3)(0,1.5){6}{\line(5,-3){0.417}}
\multiput(5.25,2.75)(0,1.5){6}{\line(5,3){0.417}}
\multiput(5.67,3)(0,1.5){6}{\line(5,-3){0.417}}
\multiput(6.083,2.75)(0,1.5){6}{\line(5,3){0.417}}
\multiput(6.5,3)(0,1.5){6}{\line(5,-3){0.417}}
\multiput(6.917,2.75)(0,1.5){6}{\line(5,3){0.417}}
\multiput(7.33,3)(0,1.5){6}{\line(5,-3){0.417}}
\multiput(7.75,2.75)(0,1.5){6}{\line(5,3){0.417}}
\multiput(8.17,3)(0,1.5){6}{\line(5,-3){0.417}}
\multiput(8.583,2.75)(0,1.5){6}{\line(5,3){0.417}}
\multiput(9,3)(0,1.5){6}{\line(5,-3){0.417}}
\multiput(9.417,2.75)(0,1.5){6}{\line(5,3){0.417}}

\multiput(4,3.5)(0,1.5){6}{\line(5,3){0.417}}
\multiput(4.417,3.75)(0,1.5){6}{\line(5,-3){0.417}}
\multiput(4.83,3.5)(0,1.5){6}{\line(5,3){0.417}}
\multiput(5.25,3.75)(0,1.5){6}{\line(5,-3){0.417}}
\multiput(5.67,3.5)(0,1.5){6}{\line(5,3){0.417}}
\multiput(6.083,3.75)(0,1.5){6}{\line(5,-3){0.417}}
\multiput(6.5,3.5)(0,1.5){6}{\line(5,3){0.417}}
\multiput(6.917,3.75)(0,1.5){6}{\line(5,-3){0.417}}
\multiput(7.33,3.5)(0,1.5){6}{\line(5,3){0.417}}
\multiput(7.75,3.75)(0,1.5){6}{\line(5,-3){0.417}}
\multiput(8.17,3.5)(0,1.5){6}{\line(5,3){0.417}}
\multiput(8.583,3.75)(0,1.5){6}{\line(5,-3){0.417}}
\multiput(9,3.5)(0,1.5){6}{\line(5,3){0.417}}
\multiput(9.417,3.75)(0,1.5){6}{\line(5,-3){0.417}}

\multiput(4.417,3.75)(0,1.5){5}{\line(0,1){0.5}}
\multiput(4.83,3)(0,1.5){6}{\line(0,1){0.5}}
\multiput(5.25,3.75)(0,1.5){5}{\line(0,1){0.5}}
\multiput(5.67,3)(0,1.5){6}{\line(0,1){0.5}}
\multiput(6.083,3.75)(0,1.5){5}{\line(0,1){0.5}}
\multiput(6.5,3)(0,1.5){6}{\line(0,1){0.5}}
\multiput(6.917,3.75)(0,1.5){5}{\line(0,1){0.5}}
\multiput(7.33,3)(0,1.5){6}{\line(0,1){0.5}}
\multiput(7.75,3.75)(0,1.5){5}{\line(0,1){0.5}}
\multiput(8.17,3)(0,1.5){6}{\line(0,1){0.5}}
\multiput(8.583,3.75)(0,1.5){5}{\line(0,1){0.5}}
\multiput(9,3)(0,1.5){6}{\line(0,1){0.5}}
\multiput(9.417,3.75)(0,1.5){5}{\line(0,1){0.5}}
\multiput(9.83,3)(0,1.5){6}{\line(0,1){0.5}}

\put(4,7){\line(5,3){5.417}}
\put(4,7){\line(5,-3){5.417}}

\thicklines
\put(4,7){\line(5,1){7}}
\put(4,7){\line(5,-1){7}}
\put(4,7){\line(-1,4){1}}
\put(4,7){\line(-1,-4){1}}
\thinlines
\multiput(4,7)(0.5,0.1){10}{\line(1,-1){4}}
\multiput(4,7)(-0.125,-0.5){7}{\line(1,-1){4}}
\multiput(4,7)(0.5,-0.1){10}{\line(1,1){4}}
\multiput(4,7)(-0.125,0.5){7}{\line(1,1){4}}

%

%
%5

\end{picture}
\caption{The conical singularity at the
origin. The two different shaded regions represent two local
coordinate systems needed to cover the tetrahedron vertex.}
\end{figure}


\newpage
\begin{thebibliography}{99}

\bibitem{bol} H. W. Kroto, J. R. Heath, S. C. O'Brien, R. F. Curl
and R. E. Smalley, Nature {\bf 318} (1985) 162.

\bibitem{ca} A nice description of the fullerenes and their
most curious features can be found in R. F. Curl and S Smalley,
Sci. Am. {\bf 265} (1991) 54; D. R. Huffman, Physics Today
November 1991, 22 and in K. Prassides and H. Kroto,
Physics World April 1992, 44.

\bibitem{spec} R. E. Stanton and M. D. Newton, J. Phys. Chem.
{\bf 92} (1988) 2141.

\bibitem{samuel} S. Samuel, {\it On the Electronic Spectrum
of $C_{60}$}, CCNY-HEP-92/5, May 1992.

\bibitem{nos} J. Gonz\'alez, F. Guinea and M.A.H. Vozmediano,
Phys. Rev. Lett. {bf 69} (1992) 172.

\bibitem{am} See, for instance, N. W. Ashcroft and N. D. Mermin,
{\em Solid State Physics} (Holt, Rhinehart and Winston, New
York, 1976).

\bibitem{des} S. Deser, R. Jackiw and G. `t Hooft, Ann. Phys.
{\bf 152} (1984) 220.

\bibitem{col} S. Coleman, {\em The Magnetic Monopole Fifty Years
later}, in {\em The Unity of the Fundamental Interactions}
(Plenum Press, New York, 1983).

\bibitem{bd} N. D. Birrell and P. C. W. Davies, {\em Quantum
Fields in Curved Space} (Cambridge University Press, Cambridge, 1982).

\bibitem{sam} S. Samuel, {\em The Solution of the Ising Model in the
Truncated Icosahedron Lattice of $C_{60}$}, CCNY-HEP-92/3.

\bibitem{tetr} S.R. White, Phys. Rev. {\bf B45} (1992) 5062.

\bibitem{cap} J. W. Mintmire, B. I. Dunlap and C. T. White,
Phys. Rev Lett. {\bf 68} (1992) 631.

\bibitem{neg} A. L. Mackay and H. Terrones, Nature (London) {\bf
352} (1991) 762; D. Vanderbilt and J. Tersoff, Phys. Rev. Lett.
{\bf 68} (1992) 511.

\bibitem{berry} M.V.Berry, Proc. R. Soc. London {\bf A 392} (1984) 45.

\end{thebibliography}
\end{document}